\documentclass[twocolumn]{aastex631}
\usepackage{amsmath}
\usepackage{ulem}
\usepackage{xcolor}   
\usepackage{url}

\catcode`_=\active
\newcommand_[1]{\ensuremath{\sb{\mathrm{#1}}}}
\catcode`^=\active
\newcommand^[1]{\ensuremath{\sp{\mathrm{#1}}}}

\def\pjd#1{}
\newcommand{\pja}[1]{#1}

\newcommand{\figstats}
{%
\begin{figure*}
\includegraphics[width=\linewidth]{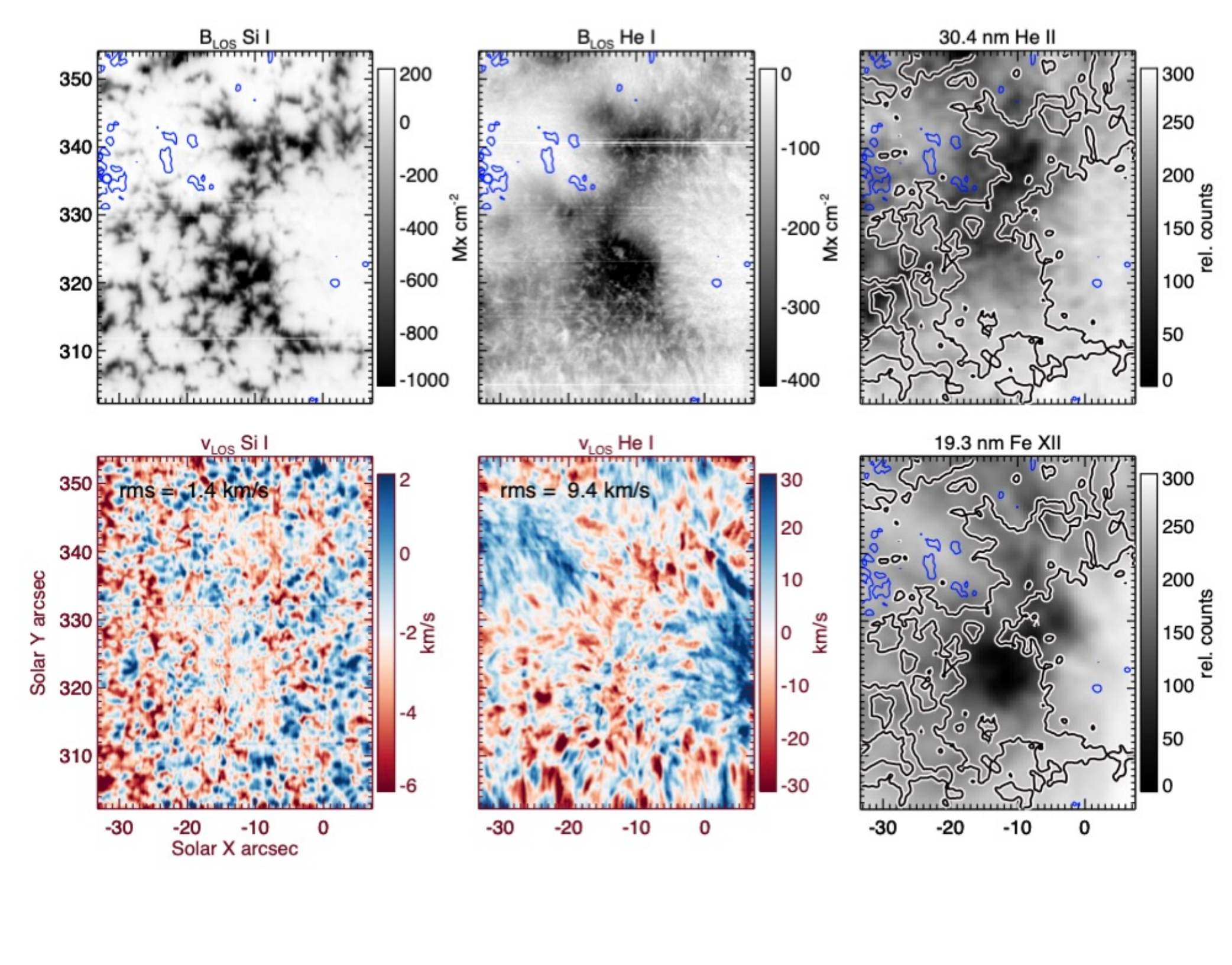} 
\caption{Images of active region NOAA 12773 on
  September 28 2020 spanning the area scanned by GRIS are shown.  
  Top
  row: LOS photospheric magnetic field, LOS
  chromospheric magnetic field, and intensity of the 30.4 nm channel of
  AIA.  Bottom row: LOS Doppler shifts of the cores of
  \ion{Si}{1} and \ion{He}{1} lines, and the intensity of the 19.3 
  nm channel of AIA.  Doppler shifts are shown in red  and blue (away from and towards the observatory).
    The color table of the AIA data is  reversed
(i.e., dark = brighter emission), superposed with photospheric GRIS B$_{LOS}$ contours at
+6, -150  Mx~cm$^{-2}$ 
in blue and black 
respectively. \pja{Blue contours are also shown in the uppermost panels.}
}
\label{fig:stats} 
\end{figure*}
}

\newcommand{\figfov}
{%
\begin{figure*}
\includegraphics[width=\linewidth]{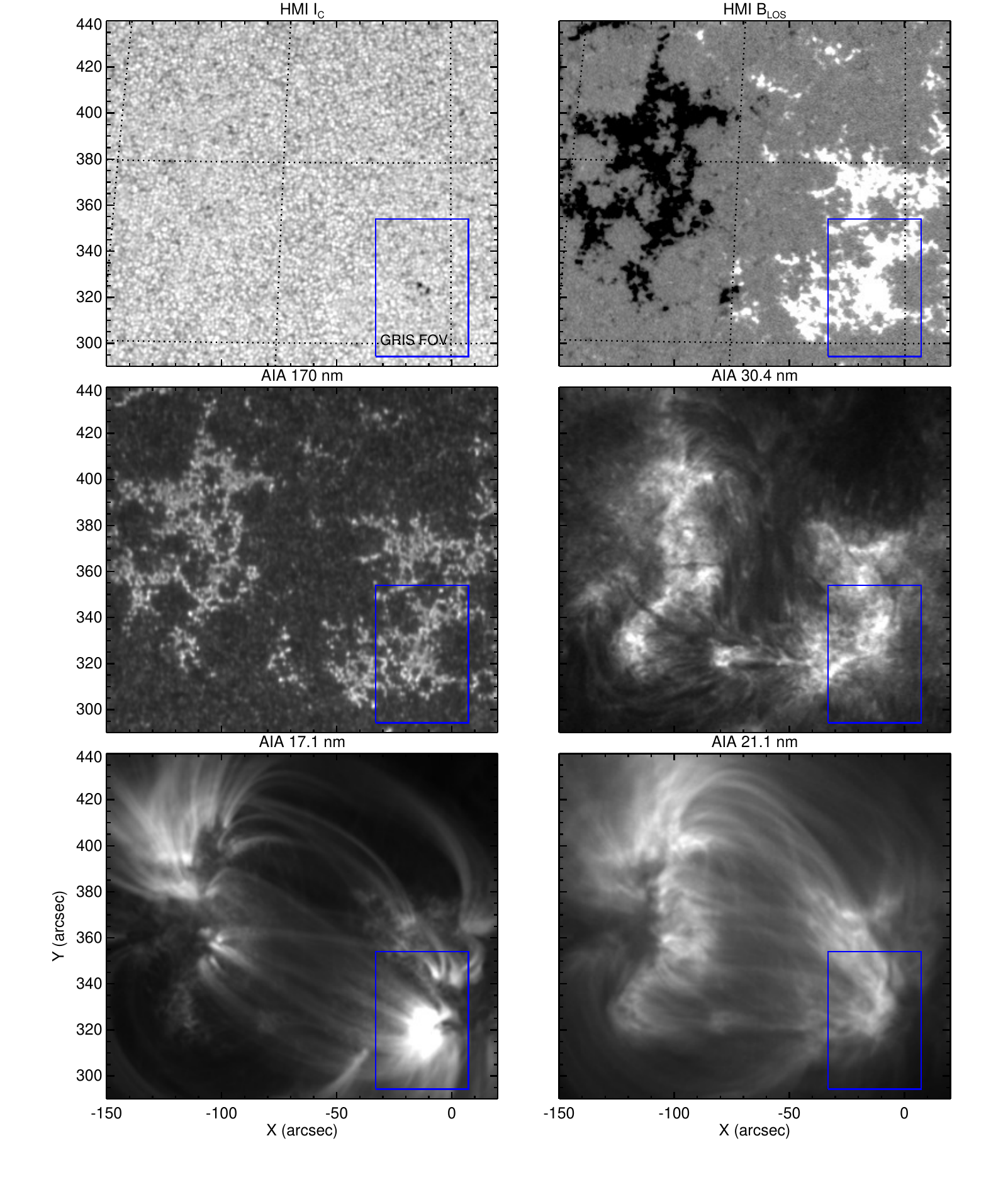} 
\vspace{-10mm}
\caption{Images are shown from the SDO spacecraft
  obtained near 19:41 UT on 28 September 2020, the mid-time of the scan with GRIS, of the active region NOAA 12773. The
  field of view scanned by the GRIS is indicated as a blue rectangle.  The slit
  was oriented 2.4 degrees clockwise from the N-S direction, and the slit scanned in the E-W direction.  North is
  upwards in the figure.
}
\label{fig:fov} 
\end{figure*}
}

\newcommand{\figha}{
\begin{figure*}
    \centering
    \includegraphics[width=\linewidth]{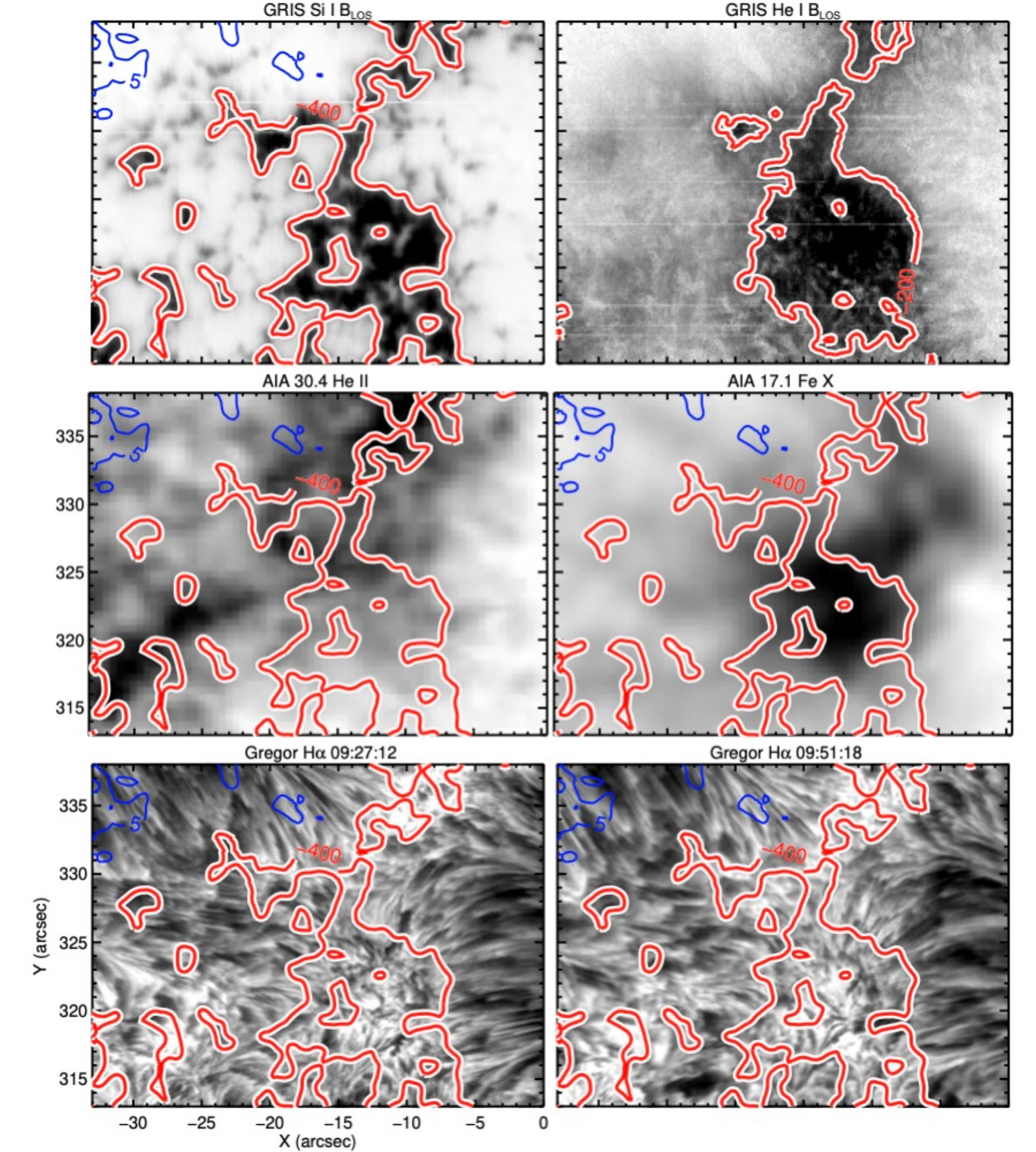}
    \caption{A close view of chromospheric magnetic field, AIA images and H$\alpha$ images from the HiFI+ instrument on GREGOR. 
    The two H$\alpha$ images, separated by 24 minutes, show a similar morphology, with fibrils frequently aligned nearly perpendicular to the -400 Mx~cm$^{-2}$ contour (red/white) of photospheric magnetic fields. }
    \label{fig:ha}
\end{figure*}
}

\newcommand{\fighamovie}{
\begin{figure}
\vskip 36pt
    \centering    \includegraphics[width=0.9\linewidth]{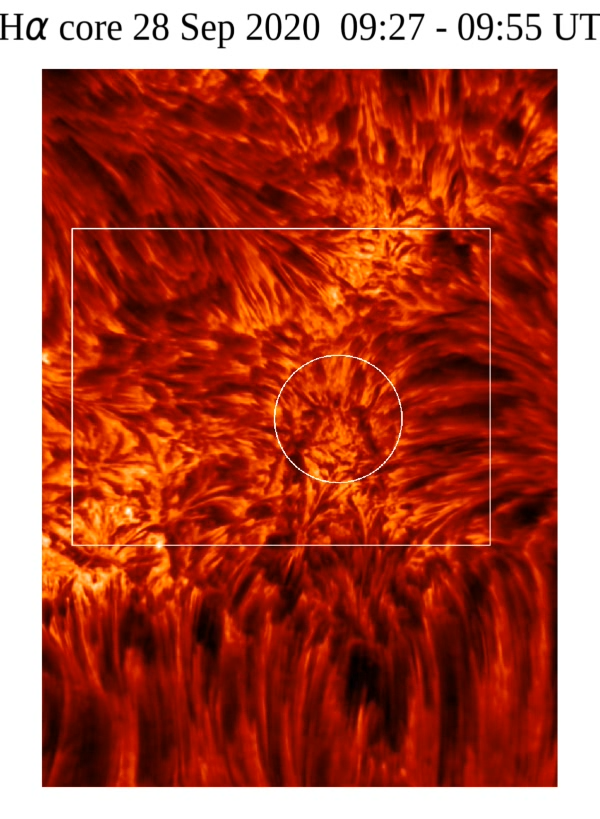}
    \caption{A snapshot of a movie of H$\alpha$ images 
    from the HiFI+ instrument on GREGOR is shown. The online movie shows a frame every 6 seconds for 28 minutes.  The
    movie shows mostly signatures of flows along
    the long striation patterns.
    The box shows 
    the area scanned by the GRIS instrument, and  the circle highlights the footpoint of the 
    most intense EUV coronal emission.  The movie reveals
    that rare flare-like signatures, presumably associated with reconnection, are visible only far from the footpoint area. In the text this
    is interpreted to mean there is 
    no clear indication of reconnection at visible scales at the footpoint of the bright coronal emission.
}
    \label{fig:hamovie}
\end{figure}
}

\newcommand{\figprofiles}{
\begin{figure*}
\includegraphics[width=\linewidth]{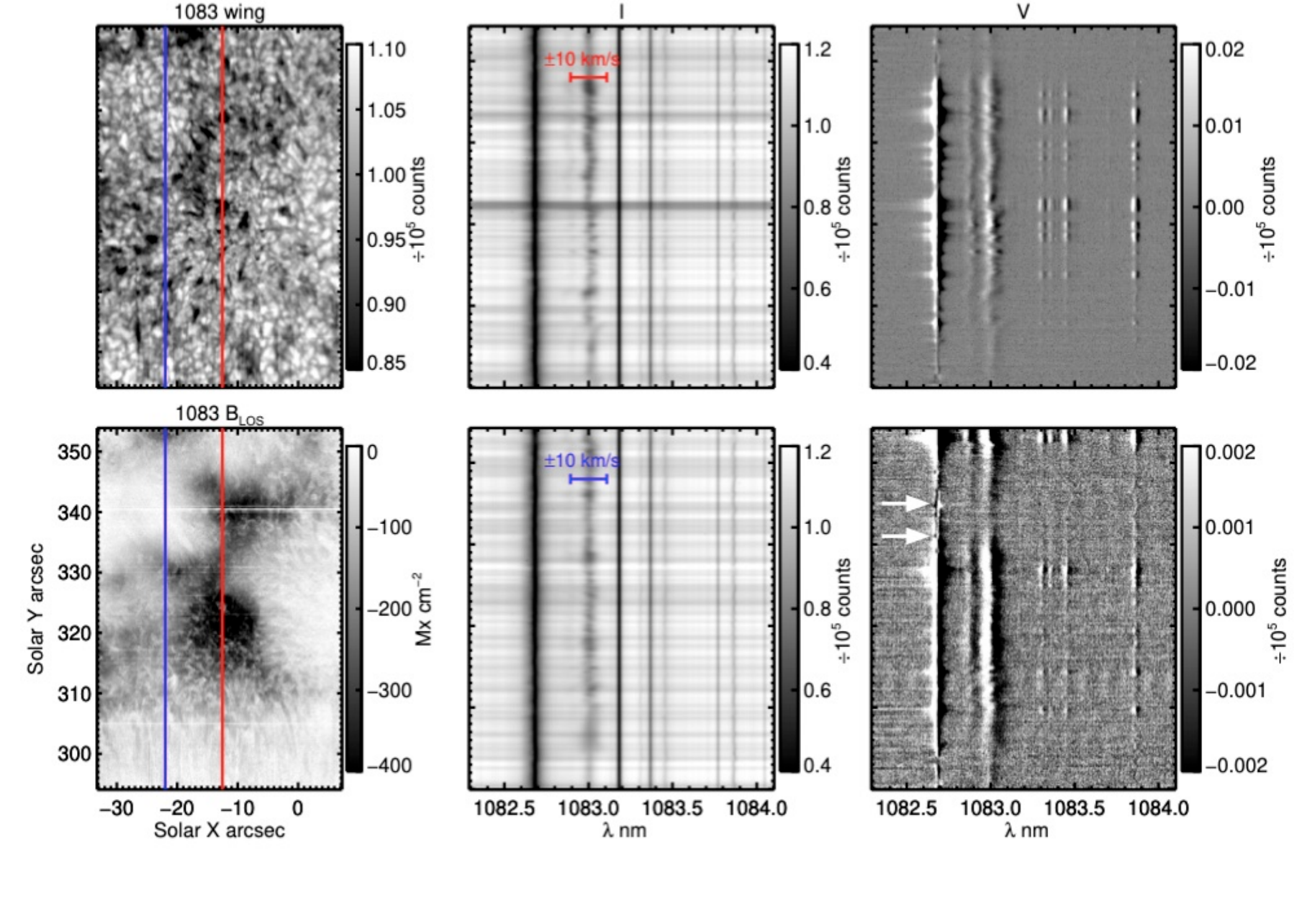} 
\vspace{-15mm}
\caption{Images show 
line wing intensity and 
derived LOS magnetic flux density 
(left panels), and spectra 
of intensity (middle panels)
and circular polarization
(right panels), from  active region NOAA 12773. The four spectra are taken from the 
red and blue vertical slit positions shown in the left panels, the upper rows are from the red position, the lower from the blue.   The $I$ and $V$ images are in units of
$10^5$ counts.  
The Stokes $V$ profiles 
in the bottom right panel show two regions
of minority polarity in the \ion{Si}{1} 
1082.7 nm line beneath the 
majority polarity 
of the \ion{He}{1}
1083 nm line, indicated by white arrows.}
\label{fig:profiles} 
\end{figure*}
}

\newcommand{\figcloseup}
{%
\begin{figure*}
\includegraphics[width=\linewidth]{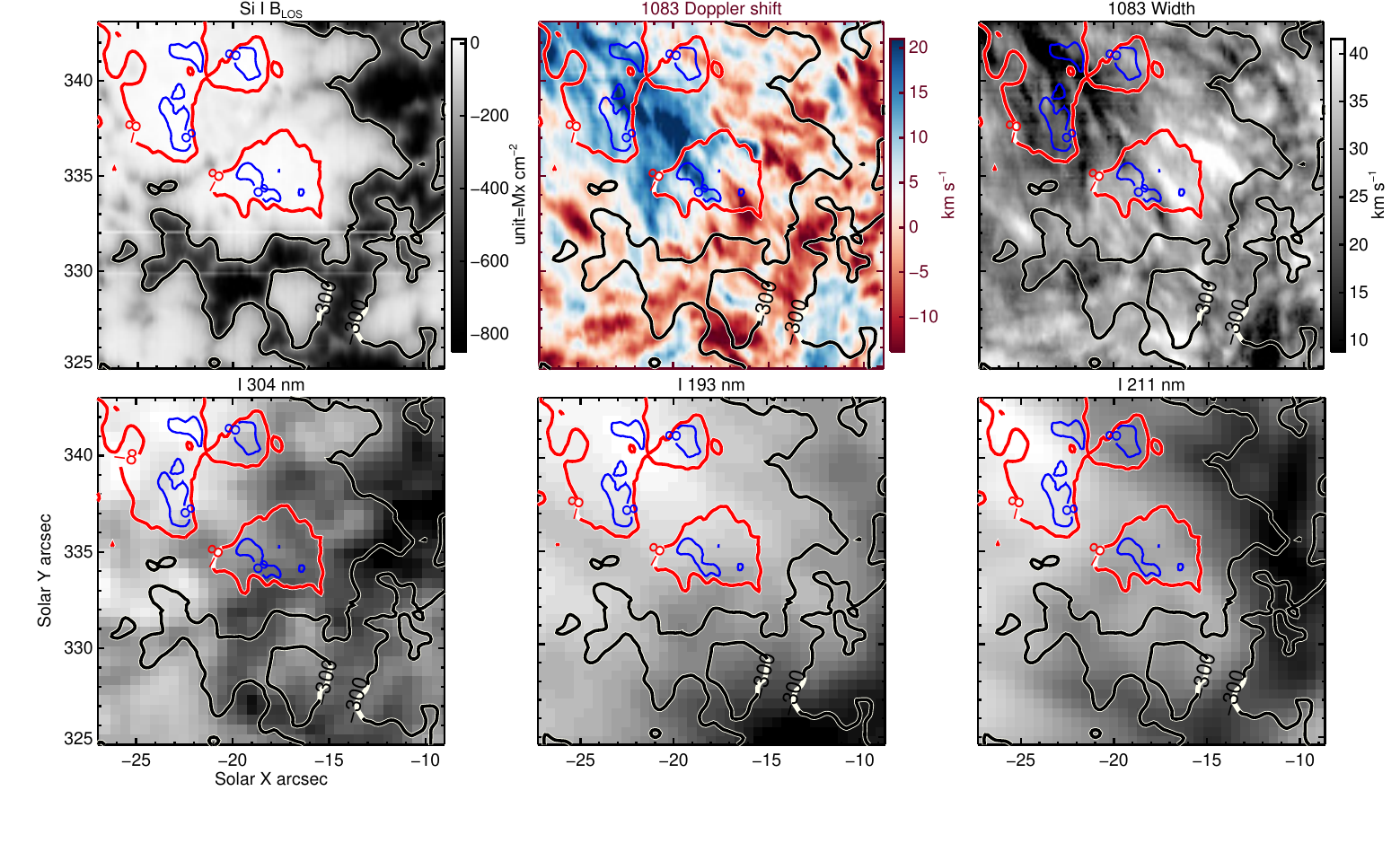} 
\caption{Minor polarity photospheric magnetic flux is highlighted within 
blue contours (8 Mx~cm$^{-2}$) together
with majority polarity contours at -8 (red) and -300 
Mx~cm$^{-2}$ (black).  In the top row 
the images show 
LOS photospheric 
magnetic flux density (left
panel), \ion{He}{1} 1083 nm  Doppler velocity density (middle panel), and width 
(right panel).   AIA data are shown in the lower three panels, darker
image density means 
brighter emission.   The 
minor polarity photospheric flux patches are encircled
by major polarity flux, and exhibit no correlations with any of the features 
formed in the upper chromosphere (\ion{He}{1}
properties} or above (AIA
brightnesses).
\label{fig:closeup} 
\end{figure*}
}

\newcommand{\hao}{
High Altitude Observatory,
National Center for Atmospheric Research,
Boulder CO 80307-3000,
 USA}

\shorttitle{}
\shortauthors{Judge et al.}

\begin{document}

\title{Magnetic fields beneath active region coronal loops}


\correspondingauthor{Philip G. Judge}
\author{Philip G. Judge} 
\affiliation{Visitor, Astronomical Institute of the University of Bern, Sidlerstrasse 5, 3012 Bern, and\\ \hao}

\author[0000-0002-7791-3241]{L. Kleint}\affiliation{University of Bern, Astronomical Institute, Sidlerstrasse 5, 3012 Bern, Switzerland}

\author[0000-0002-3242-1497]{C. Kuckein}\affiliation{ 
Max-Planck-Institut f\"ur Sonnensystemforschung, Justus-von-Liebig-Weg 3, 37077 G\"ottingen, Germany \\
   \and
   Instituto de Astrof\'isica de Canarias (IAC), V\'ia L\'actea s/n, E-38205 La Laguna, Tenerife, Spain  \\
   \and
   Departamento de Astrof\'\i sica, Universidad de La Laguna, E-38206 La Laguna, Tenerife, Spain
}


\date{Accepted. Received ; in original form }


%
%

\begin{abstract}
We examine the hypothesis 
that multipolar magnetic fields advected by photospheric granules can 
contribute heating to the  active chromosphere and corona.  On 28 September 2020 
the GRIS and HiFI+ instruments at the GREGOR telescope obtained 
data of  NOAA 12773.  We analyze
Stokes profiles of spectral lines of  \ion{Si}{1}
and \ion{He}{1}, to study 
magnetic fields from photosphere to the upper chromosphere. 
Magnetogram and EUV data from
the HMI and AIA instruments on the SDO spacecraft are co-aligned and studied in
relation to the GRIS data.  At coronal loop footpoints, 
minor polarity fields comprise just 
0.2\%{} and 0.02\%{} of the flux measured over the $40\arcsec\times60\arcsec$  area observed in the 
photosphere and upper chromosphere, centered 
$320\arcsec$ from disk center.
Significantly, the minority fields are 
situated $\gtrsim$ 12$\arcsec$ from bright footpoints. 
We use physical arguments to show that any unresolved minority flux 
cannot reach coronal footpoints adjacent to 
the upper chromosphere. Even if
it did, the most optimistic estimate of the energy released 
through chromospheric reconnection is barely sufficient to account 
for the coronal energy losses.   Further, 
dynamical changes accompanying  reconnection between uni- and multi- polar fields are 
seen neither in the \ion{He}{1} data nor in narrow-band 
movies of the H$\alpha$ line core. 
We conclude that the hypothesis must be rejected.  Bright chromospheric, transition region and coronal
loop plasmas must be heated by mechanisms involving 
unipolar fields.
\end{abstract}


\section{Introduction}
\label{sec:statement}

Many mechanisms  for driving the heating and dynamics of
corona plasma have been proposed since \citet{Edlen1943}
first proved that the plasma temperature exceeds that of the Sun's surface
by a factor of at least fifty.  The problem is multi-faceted and it has resisted
definitive solution for a variety of good physical reasons
\citep[highlighted recently by][]{Judge+Ionson2024}.  The problem can be usefully
divided into three parts: what is the nature of the source of
upwardly propagating energy?  How does the energy propagate through the atmosphere? How is this energy transformed irreversibly into heat? The purpose of the present paper
is to examine the generation of magnetic energy via the evolving 
magnetic field
immediately beneath the corona.

A significant fraction of heating mechanisms invoke the generation of
upward-directed magnetic energy through the interaction and
reconfiguration of small-scale magnetic fields which are advected by
photospheric granular flows.  These flows can generate Poynting fluxes
of electromagnetic energy both steadily, as the emerging magnetic
bundles of flux are entwined around each other
\citep{Parker1972,Parker1988,Pontin+Hornig2020}, as MHD waves are generated
\citep{Osterbrock1961,Uchida+Kaburaki1974,Hollweg1978,Howson+others2020}, and as magnetic reconnection beneath the corona causes
bulk fluid flows, small flares 
and related \pjd{physical} phenomena \citep[e.g.][]{Sturrock1968,Shibata+others2007, Nelson+others2019,Carlsson+others2019}.  \pjd{All of these
pictures are sources of magnetic free energy, and all have been
proposed to be responsible for coronal heating
\citep{Kuperus+Ionson+Spicer1981,
Zirker1993,
Klimchuk2006,
DeMoortel+Parnell2015,
Longcope+Tarr2015,
Judge+Ionson2024}. }

 Proposals involving 
magnetic reconnection of rapidly evolving granular fields
\citep{Lin+Rimmele1999}
with long-lived supergranular 
network boundary fields \citep{Leighton1959, Leighton+Noyes+Simon1962} 
have formed the basis 
of models of coronal heating for several decades.
\pjd{
\citep[e.g.][]{Title+Schrijver1998,Schrijver+others1998,Priest+others2002,Wang2016,Chitta+others2017,Chitta+others2023}. 
}
\citet{Title+Schrijver1998} suggested that
magnetic reconnection driven 
by granular convection, observed 
as a ``magnetic carpet'' of continuously
evolving photospheric fields being replaced about once per day, naturally leads to 
overlying coronal heating.  
The idea continues to stimulate later work 
\citep[e.g.][]{Schrijver+others1998,Priest+others2002,Wang2016,Chitta+others2017,Chitta+others2023}.
\pjd{All of these ideas have been
examined using observations, with the implicit assumption that
those processes responsible for heating the corona can be measured
on observable scales.  It is unclear if this assumption is valid,
even though it is almost universally adopted, for physical reasons 
discussed at some length by \citet{Judge+Ionson2024}.  The diffusion of
magnetic fields during 
magnetic ``reconnection'' occurs on unobservably small scales in 
highly conducting plasma like the corona.  Accordingly it can have 
macroscopic and microscopic effects, such as are regularly observed 
during flare and CME dynamics on the one hand, and hypothesized 
during the intricate dynamics of 
MHD turbulence on the other \citep[e.g.][]{Schekochihin2022}. }
A 
search for the term ``magnetic carpet'' 
along with ``corona'' in the NASA ADS database reveals over 60 diverse
abstracts using this concept. 
Related is the concept of
``interchange reconnection'',
frequently advocated to explain both the evolution of
coronal hole boundaries, and 
as a power source for 
the solar wind
\citep{Nash+others1988,Fisk2003}.   
In all such studies, 
a direct connection is made
between observed mixed polarity photospheric
magnetic fields and heating of the  corona. 
Here we examine this picture for active region loops \pja{which result from 
the most intense form of coronal heating.}
While magnetic fields at loop footpoints are locally 
almost unipolar \citep{Giovanelli1980}, the magnetic carpet picture \pja{nevertheless} continues to
be considered viable there \citep{Priest+others2002,Priest+others2018,
Chitta+others2017,
Chitta+others2023}.

\figfov

The primary difference of the present article 
with earlier work is 
in use of measurements of chromospheric magnetic and velocity fields
to assess coronal heating mechanisms.  
Chromospheric heating has recently been analyzed 
using  GREGOR data similar to those analyzed here  
by \citet{Anan+others2021}.

\section{Observations}

Here we examine datasets obtained using the 
GRIS instrument 
\citep{Collados+others2012}
from 
the GREGOR telescope
\citep{GREGOR2012,Kleint+others2020} 
augmented by narrow-band 
images in the H$\alpha$ line core and continuum from the HiFI+ instrument 
\citep{HIFI2023}.
Standard products from HMI  and AIA were used
to define the frame of reference and to compare with images of
the heated plasma above. 
The field-of-view observed by the GRIS instrument  is shown as a rectangle in
Figure~\ref{fig:fov}.
GRIS is a scanning slit spectropolarimeter, which 
 builds up an image of the solar spectrum
in the four Stokes parameters $I,Q,U$ and $V$
by letting light pass through a slit to a diffraction grating.
The dispersed spectra were focused for each polarization state on a 2D detector, with
wavelength in the dispersion direction, and position along the 1D slit perpendicular to this.  The slit was moved in 300 steps of $0\farcs135$  across the solar image
to build up a scan over 29 minutes and 30 seconds with a field of view (FOV) of 40.5 x 59.8 arcsec.   Examples
of the level-1 data acquired are
shown in Figure~\ref{fig:profiles}.
\figprofiles

\begin{table*}
\caption{GREGOR observations examined from September 28 2020 }
\centering
\begin{tabular}{lcllccccccl}
  \hline\hline
Instrument  & Mode & Start & End & Cadence &FOV &spatial & detector & $\lambda$ &$\Delta\lambda$&\\
& & \multicolumn{2}{c}{UT} & s& arcsec & pixel\arcsec& format &\multicolumn{2}{c}{nm} & \\
  \hline
  GRIS & IQUV &09:27:01 & 09:56:24 & 5.9& $40.5^\ast \times 59.8$& 0.135 & $443 \times 1010$ &1082.322-1084.136 & 0.018& \\
   &  & & & & & &
   (solar Y $\times\ \lambda$)
   &
  & \\
  HiFI+ No.2 & I & 08:26:38
 & 10:34:33& 6& $76.5 \times 60.5$ & 0.0498 &$1536 \times 1216$ & 656.279 & 0.060\\
  \hline
\end{tabular} 
\\
$^\ast$Slit is $59\farcs8$ long, $40\farcs5$ equals 300 steps of $0\farcs135$ in solar X (E-W direction).\hfill  
\label{tab:obs} 
\end{table*}

When reduced to a level-1 data product, data cubes of the form
\begin{equation}
S(x,y,\lambda,i),
\end{equation}
are constructed, functions of the two spatial pixels $(x,y)$ on the Sun, wavelength $\lambda$ and
one of the four polarization states labeled with index $i$.

\begin{table}
\caption{Statistics of LOS magnetic fluxes within the region scanned by GRIS}
\begin{center}
\begin{tabular}{llllll}
  \hline\hline
  $\lambda$ & Ion& Instr. & 
--ve flux & +ve flux & +ve $\div$\\
nm &          & & Mx & Mx &  --ve \\
\hline
617.3 & \ion{Fe}{1} & HMI$^\ast$&
   -1.2(21)$^\dag$  & 7.6(18)&   -0.006 \\
1082.7 & \ion{Si}{1} & GRIS& -2.0(21) &    4.0(18)   & -0.002\\
1083.0 & \ion{He}{1} & GRIS& -1.1(21) &    2.6(17)   & -0.0002\\
  \hline
\end{tabular} 
\end{center}
$^\ast$The 45s data product was used in this article. 
$^\dag$The notation is such that $7.6(18)\equiv 7.6\times10^{18}$.
The area scanned by GRIS, analyzed in common with HMI,  is
$1.3\times10^{19}$ cm$^2$. The flux sensitivities
of the measurements of the three lines listed 
are $3\times 10^{16}$, $3\times 10^{14}$, and $ 10^{15}$ Mx respectively. 
\label{tab:flux} 
\end{table}

\subsection{GRIS observations from September 28 2020}

Table~\ref{tab:obs} lists
 properties of the two
GREGOR instruments with data analyzed here.  
From Figure~\ref{fig:fov}, the HMI data reveal that GRIS scanned just one major polarity
of the active region NOAA 12773, the strongest then on the solar disk.        
Figure~\ref{fig:stats} displays line-of-sight (``LOS") magnetic flux densities and
velocities, both along the LOS, for the photospheric
line of \ion{Si}{1} (left) and chromospheric line of \ion{He}{1}
(center) respectively.  
These were derived using  same procedures 
 as in the paper by \citet{Judge+others2024}.
The  magnetic LOS fluxes, derived from
the weak field 
approximation applied to the $I$ and $V$ Stokes 
profiles of both lines, are listed in Table~\ref{tab:flux}.  \pja{The effective 
Land\'e g-factors used for 1082.7 and 1083.0 lines of \ion{Si}{1}
and \ion{He}{1} were 1.58 and 1.25 for longitudinal Zeeman components, 
respectively, and 2.45 and 1.53 for the transverse components. 
}

The magnetic sensitivity of HMI to LOS fields is roughly $\sigma(B_{LOS}) \equiv 10$ Mx~cm$^{-2}$, which gives 
$2.6\times 10^{16}$  Mx for each 0.504\arcsec$\times$0.504\arcsec{} pixel \citep{Liu+others2012}. In contrast, that 
of 
the GRIS observations for the 1082.7 nm line of 
\ion{Si}{1} is $ \sigma(B _{LOS})\sim 3.3$ Mx~cm$^{-2}$ measured 
from high frequency noise of the derived $B_{LOS}$ data
themselves. The projected GRIS pixel areas are 0.135\arcsec$\times$0.135\arcsec{}, leading to
a sensitivity of LOS photospheric fluxes of $\sim 3\times 10^{14}$ Mx.   For the chromospheric \ion{He}{1} line, the sensitivities were measured to be  three times larger than measured for the \ion{Si}{1} line. 

Figure~\ref{fig:stats} displays two contours of LOS magnetic flux density,
the blue contour traces a value of +6 Mx~cm$^{-2}$, the black contour
a value of -150 Mx~cm$^{-2}$.   Taken together with Figure~\ref{fig:fov},
it is evident that GRIS measured magnetic properties at the footpoints of a group of 
coronal loops, themselves  extending across this active region roughly from 
$X=0,Y=320$ to $X=-100$ and $Y \approx 350-410$.   
\figstats
From Figure~\ref{fig:stats}
and Table~\ref{tab:flux}
we note, in particular:
\begin{itemize}
    \item GRIS detects \pja{$1.7\times$ more photospheric 
    LOS flux than HMI. We believe this
    difference reflects the higher
    sensitivity and angular resolution  achieved by GRIS during the raster scan.}
    \item Less than 0.2\%
    of the photospheric magnetic flux covering this area is of opposite polarity.
    \item But less than 0.02\%
    of the chromospheric magnetic flux is of minority polarity.      
    \item Doppler shifts of the \ion{He}{1} line core have amplitudes  typically $\lesssim$ 10 km~s$^{-1}$ (Figures~\ref{fig:profiles} and \ref{fig:stats}).
\end{itemize}

\subsection{HiFI+ data}

Simultaneously with the GRIS
observations, 
narrow-band images of the core of the H$\alpha$ lines at 656.3 nm 
were obtained using HiFI+ 
(Table~\ref{tab:obs}). These
data were acquired in rapid  burts, 
reduced with the software package {\tt{sTools}} \citep{Kuckein+others2017}
and reconstructed using
the multi object multi-frame deconvolution algorithm (MOMFBD, \citealp{MOMFBD}), and co-aligned 
to sub-pixel precision.   The
average of all such images between 09:27 and 09:55 UT were then co-aligned with the 
\ion{He}{1} GRIS images of
line core intensity, line width and Doppler shift.  H$\alpha$ core intensity images are naturally quite different from
these GRIS images.  However we
are certain of this multi-instrument co-alignment 
to a precision of $\pm1\arcsec$ 
because there are several
features in common in both sets of images. Two such 
images can be found 
below (Figures~\ref{fig:closeup} and 
\ref{fig:ha}). 

\section{Analysis}

The above data are well matched to test the hypothesis that multipolar photospheric magnetic fields
influence coronal heating.  The first challenge to this idea is the dramatic imbalance between major
and minor polarity photospheric fluxes. Similar to five other areas 
observed with the ViSP instrument on the 
DKIST \citep{Judge+others2024,Judge+Kuin2024}, 
the minor polarity flux measured is just 0.2\%{} of the entire flux.  In the chromosphere this fraction 
drops by a factor of 10. In an appendix, \citet{Judge+Kuin2024} present physical 
arguments
as to why opposite polarity flux cannot simply be hidden as a result of 
cancellation of $V$ signals in the low $\beta$
environment of the upper chromosphere.
The second important point is that minor polarity fields should
be found geometrically close 
to footpoints of coronal plasma loops.
From Figure~\ref{fig:stats} (bottom  right panel) we see that the opposite 
polarity flux lies about 10 Mm from the nearest bright coronal plasma. 
The white arrow 
in Figure~\ref{fig:profiles}
indicates clearly 
a minority polarity 
signature in the photosphere that is
absent in the chromospheric data.  Opposite polarity fields
do not reach the chromospheric formation height of \ion{He}{1}
1083 nm. Further, the chromospheric ``fibrils'' clearly traced by 
the Doppler shifts 
of the \ion{He}{1}
lines (below $Y=303$ in the lower middle panel
of Figure~\ref{fig:stats})
clearly lie above multiple concentrations of minority polarity flux.

The image of
$B_{LOS}$ of the \ion{He}{1} line contains 
several kinds of structures at the resolution of $\approx 200$ km: the dense cores
of strong fields appear 
uniform.  Around these 
are mottled region (see the gray-white patterns
near $Y=310\arcsec$
in Figure~\ref{fig:profiles}).  Then there are 
signatures of chromospheric fibrils 
most readily seen at $Y < 307\arcsec$.   These 
regions are all unipolar.
The mottle pattern
has a characteristic scale of about $2\arcsec \equiv 1.5 $ Mm.  There
is, however, no significant correlation 
with granulation patterns in the continuum observed simultaneously with GRIS
(Kendall's rank correlation coefficient $\tau$ is just 0.16 with a probability of correlation of zero).

Are these minor polarity fields
important for forming the transition region?   Supporting evidence 
would  suggest the presence of ``cool loops'' 
as a basic structure for
the transition region ]\citep[e.g.][]{Hansteen+others2014}, 
extended to these otherwise unipolar regions. 
The upper right panel of the figure 
shows that, if anything, the largest opposite polarity patches of flux lie where the 
30.4 nm line of \ion{He}{2} is weak.  Other areas of strong \ion{He}{2} emission are associated far more with the unipolar 
dominant field than anything related to the 
minority flux.

\subsection{Consequences of 
reconnection of opposite polarity fields with the magnetic network}

If significant, reconnection between granular and network fields
must 
be accompanied 
by observable plasma motions somewhere within chromospheric plasma.   
Reconnection is intrinsically faster as the Alfv\'en speed 
increases, and so is the 
subsequent plasma dynamics.
Hence any reconnection
outflows driven by newly unbalanced 
Maxwell stresses 
should be observed mostly in the upper chromosphere
where the plasma inertia is small. Changes
in magnetic connectivity manifested through fibril structures 
should also be observed as a consequence of the 
proposed mechanism, as the system
seeks a lower energy configuration after reconnection
\citep[e.g.][]{Kulsrud2011}.   The central region  of Figure~\ref{fig:stats} hosts the brightest coronal emission. This region is completely unipolar as detected by GRIS.  The LOS Doppler shifts
are unexceptional,
rms speeds on the order of the chromospheric sound speed $c_S \lesssim 10$ km~s$^{-1}$
in the regions under bright
coronal emission.  
In the photosphere,
where $\rho \approx 10^{-7}$ g~cm$^{-3}$
and $B \gtrsim 300$ G
(measured in the \ion{Si}{1} line), the Alfv\'en speed
is $c_A \gtrsim 3$ km~s$^{-1}$.  Reconnection in the 
photosphere is very slow.  
But through
the  chromosphere, $\rho$ decreases exponentially with height to $ \sim 10^{-13}$
g~cm$^{-3}$. With $B\sim 200$ G, estimated from the \ion{He}{1} 1083 nm line, $c_A$ exceeds
800 km~s$^{-1}$. There is no evidence 
of speeds remotely approaching 
such values in the regions below coronal
footpoints in the figures here or those in \citet{Judge+others2024}.  Observed
chromospheric rms speeds are almost two orders of magnitude smaller
(Figure~\ref{fig:stats}). 

We have searched for morphological changes in fibrils in H$\alpha$ data from 
the HiFI+ instrument
(Table~\ref{tab:obs}).  Figures~\ref{fig:ha} and \ref{fig:hamovie} 
show snapshots and a 6 second cadence movie of the region directly
beneath the bright 
 coronal emission.  The movie is included as online material. 
 These data
reveal that any on-going reconnection on
observable scales lies below the detection limit
of changes in magnetic fields and fluid motions, both 
Doppler and proper motion signatures. 

The qualitatively similar observations of \citet{DiazBaso+others2021} were of an 
initially unipolar pore, but with a clearly observable emerging region 
of opposite polarity and spanning several arcseconds 
nearby. Over several hours 
they measured a reduction in flux of $4 \times 10^{19}$ Mx of both polarities.  These fluxes
are an order of magnitude higher than the minor polarity flux in
the far more unipolar region reported here (Table~\ref{tab:flux}).
This flux ``cancellation'' (a descriptive not a physical term) 
was accompanied 
by ongoing bright chromospheric emission, some  coronal
emission, increasing transverse magnetic fields, falling longitudinal
fields, and images of plasmoids moving with speeds of $\sim 10^2$ km~s$^{-1}$.
None of these phenomena 
are present in the data analyzed here.   

\figcloseup

\subsection{Hidden minority fields in otherwise unipolar regions}
\label{sec:hidden}
Figures \ref{fig:stats}--\ref{fig:ha} highlight the resolved minority  polarity regions using blue contours for small values of $|B_{LOS}|$, the other contours show regions of dominant polarity at
much larger values of $|B_{LOS}|$.   The small areas of fields of minority
polarity are observed away from the dominant polarity
network regions by at least 2\arcsec. 

Some have argued that
minority-polarity 
features, if small, can 
lie undetected \textit{within or adjacent to}  network patches of  
dominant polarity, in the photosphere.
Here we examine the consequences of the proposition that such fields extend
into the \textit{chromosphere} and above.  

For  decades, this effect has been recognized in 
photospheric measurements
\citep{Stenflo1973}. Unresolved structure contributing to
the net Stokes $V$ profiles are 
described by a ``filling factor''
of the dominant unresolved magnetic field. But in the upper 
chromosphere the magnetic filling factor is close to one. We estimate the plasma $\beta$ to be $\lesssim 0.001$ where the \ion{He}{1}
lines form \citep{Avrett+Fontenla+Loeser1994}, thus
no force can 
counteract   an imbalanced  Lorentz force.  The plasma is almost force-free, filling the volume with magnetic field.  
If the chromospheric plasma were not force-free, it
would evolve at the Alfv\'en speed to make it so.
Any minority flux concentration
must therefore also have $B\sim 200$ G, the pressure being 
constant across any mis-aligned vector magnetic fields.
To be absent from
the chromospheric measurements of $B_{LOS}$
 presented here, it must lie below
 a $3\sigma$ sensitivity 
$\approx 3\times 10^{15}$ Mx. With 
$B\sim 200 $ G, the minor polarity flux must occupy an
area $A \lesssim 5\times 10^{12}$ cm$^2$, equivalent to
a square of 22 km on a side, subtending 
an angle of $0\farcs03$ at the Sun, four times smaller
than the GRIS pixel size.   We can estimate the  energy released by the complete annihilation of these fields with the network field. 
The geometry is represented by the cartoon shown as Figure~11 of \citet{DiazBaso+others2021}.
When this field is advected to the network at speed 
$u \sim 0.5$  km~s$^{-1}$, 
the total power entering the diffusion region from both sides is 
\begin{equation} \label{eq:fluxest}
    2 \frac{B^2}{8\pi } \frac{\ell u}{\sqrt{A}} \approx
     10^{9}  \ \ \mathrm{erg~cm^{-2}~s^{-1}}
\end{equation}
over the vertical area $A$, where we have taken $\sqrt{A}$ 
as the horizontal transverse size of the 
advected flux bundle and  $\ell\sim 150$ km
as the vertical size of
the reconnecting diffusion region.  For $\ell$ we use 
the  scale height at the top
of the chromosphere
where reconnection is fastest.  This energy flux density can 
be compared with $1.6\times10^8$ 
erg~cm$^{-2}$~s$^{-1}$ estimated for
the energy released 
from a much larger region of the chromosphere
via reconnection by \citealp{DiazBaso+others2021}. 
Clearly equation~(\ref{eq:fluxest}) is
a gross overestimate, the
maximum
available power in reconnection which must be averaged over 
an entire GRIS pixel
of $0\farcs135$ on a side, a horizontal area twenty times larger.   Further, 
initially just one half of the energy of reconnection travels upwards, 
and of this, at most half ultimately will
end up as plasma heating (reconnection leads initially to  
bulk acceleration, only consequent processes lead to irreversible heating). 
Taking these factors into account, an 
upper limit to
the flux density 
of energy available for heating per unit area is
$\lesssim 1.3 \times 10^7$ erg~cm$^{-2}$~s$^{-1}$. This  is similar to the required 
flux density of
$10^7$ erg~cm$^{-2}$~s$^{-1}$
\citep{Withbroe+Noyes1977}.  The proposal is tenable in terms of energy,
but only barely, and even then 
it requires fully half of the available upward directed energy
to be converted into heat.  Any misalignment of reconnected fields from anti-parallel reduces this estimate further.

Lastly, in their appendix, \citet{Judge+Kuin2024}
argued independently 
that such fields would be incompatible 
with  observations and force balance in the low plasma-$\beta$
upper chromosphere. 

\figha

\section{Conclusions}

In summary, nothing about the dataset
analyzed suggests a role for the 
small-scale multipolar granular fields in
coronal heating, contrary to the 
findings of papers based upon photospheric magnetic field
measurements
\citep{Priest+others2002,Priest+others2018,Chitta+others2017,Chitta+others2023},
but in agreement with our work using different 
chromospheric magnetic field measurements
\citep{Judge+others2024,Judge+Kuin2024}.
The differences can perhaps 
be reconciled because chromospheric measurements are 
superior to photospheric measurements for investigating 
conditions at the coronal base.  
Our work is magnetically far more sensitive than earlier work, for instance with MDI 
\citep{Schrijver+others1998}
or HMI \citep{Wang2016} instruments.  The diagnostic \ion{He}{1} multiplet cannot form
deeper than the uppermost scale height of
the chromosphere, it either forms 
there via the penetration of EUV photons 
from the hotter plasma above, or it is essentially absent 
\citep{Avrett+Fontenla+Loeser1994,Leenaarts+others2016}.    It seems to us that earlier work based on photospheric magnetic fields, including those using sophisticated 
extrapolation schemes, simply cannot 
offer the kind of penetrative diagnostics 
available to ground-based polarimetry of chromospheric plasmas.

Signatures of magnetic reconnection between observed minor polarity fields 
(blue contours in the figures),  and the dominant fields  
are simply not observed. Together with 
 physical arguments against multipolar 
chromospheric fields above plages (section~\ref{sec:hidden}), we suggest that
reconnection in chromospheric plasma directly beneath typical active region loop footpoints does not play a role in heating the loops. 
We conclude that, along with a study of
the 854.2 nm line of \ion{Ca}{2}
observed with DKIST reported by
\citet{Judge+others2024}, there is less and less room available to believe that 
opposite polarity fields have a role to
play in heating the active corona.  
This conclusion calls into question 
the role of the ``magnetic carpet''
in the heating of active coronal loops, \citep{Title+Schrijver1998,Priest+others2002}.  This refutation, if confirmed, marks a significant advance in the quest to identify heating mechanisms by eliminating an entire class of energy sources for  commonly observed  plasma loops \citep{Judge+Ionson2024}.

It will
be interesting to explore the kinds of observations presented here with other datasets, especially in regions of quiet Sun
where, although the \ion{He}{1} lines are weak, they can still be used to probe 
conditions at the coronal base \citep{Harvey+Hall1971,Ruedi+Keller+Solanki1996,Lagg+others2004}. 
In the case that the \ion{He}{1} lines are too weak to explore at sufficiently 
high angular resolution, one might use the 854.2 nm line of \ion{Ca}{2}, or
if selected for flight, various lines of \ion{Mg}{2} and \ion{Fe}{2}
in the UV region (259-281 nm) using the Chromospheric Magnetism Explorer 
instrument (\citealp{Gilbert2023}, see \citealp{Judge+others2021,Judge+others2022} for
reasons to observe in this region).

Based upon the
observations presented here and elsewhere 
\citep{Judge+others2024,Judge+Kuin2024}, we wish to make a final remark on the well-known 
dynamics and thermal fine structure in the chromosphere.    In this context, 
 we are struck by the fact that the dynamics appear to reflect mostly field-aligned motions 
 (see Figure~\ref{fig:hamovie}), 
 as 
changing pressures 
associated with granulation 
drive fluid into the chromosphere, which mostly 
returns downwards. Such motions do little  work 
on magnetic fields. Instead it is likely that much of
this eye-catching dynamics is associated
with pressure-driven sonic 
flows \citep[e.g.][who studied such motions in the context of ``spicules of type I'']{Hansteen+others2006}.  They are highly visible
because of associated 
variations in density and 
temperature.  Yet 
such motions cannot 
significantly 
influence coronal heating, 
through the same 
argument used by \citet{Athay+White1978}
to deny acoustic waves
a role in heating the 
corona.  In essence, 
unless there is clear evidence for  reconnections 
and/or cross-field 
plasma acceleration driven by flux emergence, 
the dynamic fine structure of the chromosphere is of 
no real interest for
heating the corona.  The more
supersonic spicules
of ``type II'' could in principle supply kinetic 
energy into the corona
\citep{dePontieu+others2007}, but perhaps 
the area covered by them 
 too small to explain the rather ``thick'' coronal
 loops and quieter regions which seem to be 
 larger across the dominant direction of the magnetic field
 \citep[even the thinnest coronal structures yet found seem to
exceed 250 km,][]{Rachmeler+2022}. The
product of their 
number times surface area (in the traditional interpretation
of thin, straw-like structures) is far smaller
than the area of the corona,
and given the frozen-field
condition, so it is difficult to see how they might transfer
energy across magnetic fields to fill the larger 
coronal volume.   It may be useful to keep this observation in mind when 
discussing coronal heating
and dynamics, perhaps by
trying to identify where
the chromospheric dynamics 
does work on the magnetic field extending into the corona.

\section*{Acknowledgments}

We thank the Swiss National Science Foundation 
(grant No. 216870) which made this work 
possible.
This
material is based upon work supported by the National Center for
Atmospheric Research, which is a major facility sponsored by the
National Science Foundation under Cooperative Agreement No. 1852977.
CK acknowledges funding from the European Union's Horizon 2020 research 
and innovation program under the Marie Skłodowska-Curie grant agreement No. 895955.

\fighamovie

\bibliographystyle{aasjournal}
\bibliography{best}

\end{document}